\documentclass[apj, chicago]{emulateapj}
\journalinfo{\textsc{The Astrophysical Journal}}
\slugcomment{Received 19 August 2019; Accepted 23 September 2019}
\usepackage{url}

\usepackage{natbib}
\usepackage{epstopdf}
\usepackage{graphicx}
\usepackage[colorlinks,
        hypertexnames=false,
        pagebackref=true,
        linkcolor=blue,    
    citecolor=blue,        
    filecolor=magenta,     
    urlcolor=blue          
]{hyperref}                

\usepackage{array,etoolbox}
\preto\tabular{\setcounter{magicrownumbers}{0}}
\newcounter{magicrownumbers}

\newcommand{\nar}{ {\it New Astronomy Reviews}}

\shorttitle{Finding the critical decay index in solar prominence eruptions}
\shortauthors{Vasantharaju et al}
\begin{document}
\title{Finding the critical decay index in solar prominence eruptions}
\author{N.~Vasantharaju$^1$, P.~Vemareddy$^2$, B.~Ravindra$^2$, and V.~H.~Doddamani$^1$}
\email{vrajuap@gmail.com}
\affil{$^1$Department of Physics, Bangalore University, Bengaluru-560 056, India}
\affil{$^2$Indian Institute of Astrophysics, Koramangala, Bengaluru-560 034, India}

\begin{abstract} 
The background field is assumed to play prime role in the erupting structures like prominences. In the flux rope models, the critical decay index ($n_c$) is a measure of  the rate at which background field intensity decreases with height over the flux rope or erupting structure. In the real observations, the critical height of the background field is  unknown, so a typical value of $n_{c}=1.5$ is adopted from the numerical studies. In this study, we determined the $n_c$ of 10 prominence eruptions (PEs). The prominence height in 3D is derived from two-perspective observations of \textit{Solar Dynamics Observatory} and \textit{Solar TErrestrial RElations Observatory}. Synoptic maps of photospheric radial magnetic field are used to construct the background field in the corona. During the eruption, the height-time curve of the sample events exhibits the slow and fast-rise phases and is fitted with the linear-cum-exponential model. From this model, the onset height of fast-rise motion is determined and is considered as the critical height for the onset of the torus-instability because the erupting structure is allowed to expand exponentially provided there is no strapping background field. Corresponding to the critical height, the $n_c$ values of our sample events are varied to be in the range of 0.8-1.3.  Additionally, the kinematic analysis suggests that the acceleration of PEs associated with flares are significantly enhanced compared to flare-less PEs. We found that the flare magnetic reconnection is the dominant contributor than the torus-instability to the acceleration process during the fast-rise phase of flare-associated PEs in low corona ($<1.3R_{\odot}$).
\end{abstract}

\section{Introduction}
\label{Intro}


Coronal mass ejections(CMEs) are considered as the most geo-effective phenomena that  happens in the solar atmosphere \citep{Kahler1992,Gosling1993}. These CMEs are frequently observed in association with filament or prominence structures \citep{Gosling1974}. Filaments or prominences  are the coronal structures that are two orders of magnitude more cooler and denser than the surrounding coronal atmosphere. When these structures are observed against the solar disk then they are referred to as ``filaments" and when they are observed at the solar limb, they are referred to as ``prominences". The solar prominence eruptions lead to ejection of large clouds of magnetized plasma observed in the lower and middle corona prior to the observations of CMEs in the upper corona. The topological structure of prominences is that they are supported by twisted magnetic field lines that wrap around an axial magnetic field called magnetic flux ropes (MFRs) such that the magnetic dips in their lower windings support the prominence plasma against gravity \citep{Kippenhahn1957,VanBallegooijen1989,Aulanier1998}. This suggests that the MFRs play a key role in solar eruptions and hence almost all CME-initiation models assume the presence of the flux rope structure \citep{Chen2011,Xie2013,Vourlidas2014}.

Observations in H$\alpha$ reveal that the filaments are visible for hours to several days before they erupt or disappear. This suggests that they are in equilibrium with the surrounding environment. Then the filament eruption is regarded as the loss of this equilibrium. In the flux rope-based models \citep{Kliem2006}, the filament equilibrium is the balance between the inward-directed magnetic tension of the external overlying field that embeds the flux rope and the outward-directed magnetic pressure between the flux rope’s axis and the photospheric boundary. From the ideal MHD point of view, this equilibrium state will be ruptured by two ways. One is exceeding twist in the flux rope leading to kink instability and other is the torus instability arising when there is a rapid decline of the background field in the direction of the expansion of the flux rope. However, the sheared arcade model assumes that the filament/sigmoid is composed of the sheared and twisted core field and the reconnection of the shear field lines forms the flux rope and the subsequent eruption \citep{Moore2001}.
 
Kink instability can initiate the rise motion of the flux rope to a height from where the flux rope eruption is driven by the torus instability \citep{Vemareddy2014}. Even with exceeding critical twist, the flux rope can not lead to a successful eruption when there is strong overlying field.
Recent statistical study of  36 strong flare events \citep{Jing2018} also confirms that kink instability plays a little role in the eruption of flux ropes. Therefore, the decrease of the overlying field with height (torus instability) plays a main role in deciding whether the instability leads to a confined event or to a CME. For example, the failed/confined eruption of an active region filament occurred on 2002 May 27 \citep{Ji2003,Torok2005}. This AR filament started to rise rapidly and developed a clear helical shape. Eventually it's rise motion got terminated after reaching a height of 80 Mm due to the strong strapping field and was just accompanied by a M2 flare without CME. However, there have been reports that the rotational motion of flux rope along with torus instability play significant roles in eruptivity of an event  \citep{Song2018b,Zhou2019}.
 
The idea of torus instability was first proposed by \citet{Bateman1978} in tokamaks, and first revisited for solar eruptions by \citet{Kliem2006}. According to this instability, a current ring of major radius {\em R} is embedded in an external magnetic field. The ring experiences radially-outward ``hoop-force" due to its curvature and this force decreases in magnitude if the ring expands. If the inwardly directed Lorentz force due to the external field decreases faster with major radius R than the hoop force, the system becomes unstable. Assuming an external magnetic field $B_{\rm ex}$ $\propto$ $R^{-n}$ , the decay index {\em n} is defined as $n = -\ d log (B_{\rm ex}) / d log(R)$. It means that when the decay index of the external field is equal to or higher than the critical decay index value $n_{c}$, the system becomes unstable and by any small disturbance to the current channel initiates its outward motion uninhibitedly. \citet{Titov1999} and \citet{Torok2007} have performed numerical simulations with semi-circular flux rope embedded in external field. They found that the torus-instability occurs when the flux rope axis reaches a height where the decay index of the external field is larger than $n_{c}$.

The torus instability threshold depends on the geometry of the flux rope and is subjective to the case of particular study. For thin current distribution, the critical decay indices for straight and semi-circular current-channels are 1.0 and 1.5 respectively. For thick current-channels as expected in corona, the $n_{c}$ lies in the range 1.1$-$1.3 if the cross section increases during the eruption or 1.2$-$1.5 if the cross-section remains constant \citep{Demoulin2010}. These thresholds are derived using the current-wire models, where the equillibrium properties are determined using only the momentum equation in terms of current distribution.  On the other hand, many numerical MHD simulations (for example \citealt{Torok2007,Fan2007,Demoulin2010,Fan2010}) were conducted using the full set of MHD equations to validate the torus instability and they suggest the values of $n_{c}$ in the range 1.4$-$1.9. 

\begin{table*} [!ht]
	\caption{Details of PEs along with their critical time ($T_c$) and critical height ($H_c$) of eruptions are tabulated. Computed critical decay indices ($n_c$) at $H_c$ of  sample events with the associated GOES X-ray class flares and CME linear speed (km/s) are listed.}	
	\small\addtolength{\tabcolsep}{-0.2pt}
	\begin{center}
		\begin{tabular}{|c|c|c|c|c|c|c|c|c|}
			\hline 
			P No.   &    NOAA AR    &       Date       &       Loc     &       $T_c$     &     $H_c$ ($R_{\odot}$)   &  $n_c$  &  GOES class  &   CME speed \\ 
			\hline 
			P1      &     11164  &    20110307   &   N24W62   &   19:33    &   1.035   &   1.02   &  M3.7    &       2125	\\ 
			P2  & 11207 & 20110511  &     N20W53 &     02:16 &       1.043  &    1.31  &   B8.1       &     745     \\ 
			P3      &     11232   &  20110612  &    N08W74 &     13:34   &    1.028   &   0.80     &    --  &        493	 \\ 
			P4     &      11343 &    20111109  &     N28E41 &      13:01   &    1.055   &   0.86  &   M1.1   &         907   \\
			P5      &     11386  &   20111224  &     S16E58  &      08:17   &    1.015  &   1.01  &    C5.2    &    732\\
			P6      &     11639  &   20121226  &     S16E80  &     17:02   &    1.022   &   0.88     &    --  &            240 \\
			P7    &       11667  &   20130211  &    N22W64  &    18:49  &     1.023   &   1.05  &    B5.8    &    1161\\
			P8   & 11691 &  20130316   &   N12W65  &   13:38 &       1.073  &    1.23   &     --   &      786 \\
			P9     &       12113 &   20140708  &     N09E58 &      16:05   &    1.029   &   1.09   &  M6.5      &   900\\
			P10      &    12342  &   20150509  &    N18E77  &     01:01   &    1.037  &    1.19   &   C7.4   &    661 \\
			\hline 
		\end{tabular} 
	\end{center}
	\label{tab1}
\end{table*}

To determine the $n_{c}$ in the actual observations, one needs to have the critical height of the flux rope from where it experiences rapid rise motion and the background magnetic field. Given a model of the three-dimensional (3D) background magnetic field, the critical height is still unknown due to projection effects. In the cases of the CME eruptions near the solar disk, \citet{Torok2005} proposed to use a constant value of $n_{c}=1.5$. Using this value, many studies have derived a critical height of 42 Mm being the dividing line between the confined and eruptive events \citep{Liu2008,Vemareddy2018,Vasantharaju2018}. However, a recent numerical simulation study by \citet{Zuccarello2015} clearly showed a slightly different value of $1.4\pm0.1$ at the onset of eruption.

Further, several observational studies were also made to determine the $n_{c}$. \citet{Filippov2001} performed a statistical study of 27 quiescent prominences and found that prominences are prone to erupt when they reach a critical height where the decay index of the external field is 1. Recently, \citet{McCauley2015} studied the kinematics of 106 prominence eruptions and found that the average decay index at the onset height of fast-rise phase is 1.1. Both these studies have not employed the STEREO observations to determine the true height of prominence features which means that the determined critical heights are subjective. \citet{Filippov2013ApJ} employed different observational viewpoints provided by the twin STEREO and SDO spacecrafts to study the quiescent filament eruption and found that $n_{c}$ is ~1.0. In this framework, \citet{Zuccarello2014} studied an active region filament eruption and concluded that the filament reaches a height where the decay index is in range 1.3$-$1.5. So in general, $n_{c}$ values determined by observational studies are a bit smaller than that of simulations and this apparent difference in $n_{c}$ values is mainly due to the location where exactly the torus instability criterion is evaluated \citep{Zuccarello2016}. They showed that in simulations the $n_{c}$ is computed at the flux rope axis during the onset of  the eruption whereas in observational studies the $n_{c}$ is computed  at the apex of the prominence. Owing to importance in space-weather, the critical height and decay index have become the subject of many research studies including this article. 

Motivated by the above studies, we studied the value of the $n_{c}$ in 10 prominence eruption cases. We used simultaneous STEREO and SDO observations to derive the prominence kinematics based on the true height of the rising prominence. From the kinematics profile, the prominence eruption is characterized distinctly viz., slow-rise and fast-rise phases. Corresponding to the height at which the fast-rise motion commences, the $n_{c}$ is obtained. Further, we also paid attention to the acceleration mechanism of the flux rope.  Now it is widely accepted that the flux-rope instability triggers the prominence/CME eruption first, and then magnetic reconnection underneath provides further acceleration \citep{Lin2003,Vrsnak2016}. Past numerical studies demonstrated that both these mechanisms have comparable contributions to the prominence acceleration \citep{Chen2007a,Chen2007b}. This has been confirmed observationally by analyzing the relationship between kinematics and magnetic reconnection process during an AR filament eruption \citep{Song2015} and a quiescent filament eruption \citep{Song2018a}. Both these events are associated with X-ray flares. Nonetheless, the scenario might be different from event to event. If a good temporal correlation exists between the prominence kinematics and reconnection characteristics, then magnetic reconnection is important for the acceleration process otherwise its not \citep{Song2013}.  Recent analytical study by \citet{Vrsnak2016} showed that magnetic reconnection is dominant than MHD instability in impulsively accelerated events. In order to have more insights on the acceleration mechanism, we study more number of sample events to arrive at solid conclusion. Details of the observational data and analysis procedure are given in Section ~\ref{ObsData}. The analysis results are described in Sections ~\ref{Res}. Summary and discussions are given in Section ~\ref{summ}.

\section{Observational data, Instruments and methods}
\label{ObsData}
We selected 10 active region (AR) prominences located within the longitudinal belt of 40 to 80 degree. Note that our selection of prominence source regions are not located at the solar limb so that the computation of background magnetic field is not significantly affected by the discontinuity of magnetic field on solar limb in synoptic maps. At the same time, in order to minimize the projection effects in the determination of height of the prominence features, the regions located beyond the $\pm$40 degree longitude from central meridian are selected. These prominence events, naming P1-P10, are listed in first column of Table~\ref{tab1}. The filaments in all our sample events are lying above the polarity inversion line (PIL) of source ARs except P2, P7 and P8 filaments, which are located above the neutral region between two adjacent extended bipolar regions (EBRs). The distinct PIL or neutral region beneath the eruption will be used to compute the decay index as described in the following section~\ref{ch_di}. The second, third and fourth columns in Table~\ref{tab1} gives the basic information of events, viz., source NOAA AR number, date and location of eruptions respectively. All prominence eruptions in our sample are characterized by distinct slow- and rapid-acceleration phases. The critical time ($T_{c}$) of eruptions are defined as the onset time of fast acceleration phase of prominence eruptions. $T_{c}$ for all the sample events are listed in fifth column of Table~\ref{tab1}. The true prominence apex height from the photospheric surface at the critical time refers to critical height ($H_{c}$) and is listed in sixth column of the Table~\ref{tab1}. The decay index at the critical height of prominence eruption is termed as $n_{c}$, which are tabulated in seventh column of Table~\ref{tab1}. Geostationary Operational Environmental Satellite (GOES) provides the soft X-ray (SXR) flux integrated from the full solar disk, which are used to characterize the magnitude, onset and peak times of solar flares. Eighth column gives the associated GOES X-ray class flares and ninth column gives the CME linear speed (obtained from SOHO LASCO CME catalog: \url{https://cdaw.gsfc.nasa.gov/CME\_list/}). In the following section, we describe the procedures for computing $T_{c}$, $H_{c}$, $n_{c}$  and kinematics of our sample events.

True height of the prominence feature is determined by the tie-pointing method \citep{Thompson2009}. This method is computerized as the IDL routine \texttt{scc\_measure.pro} available in SolarSoftware (SSW) distribution. For this, we use simultaneous observations in 304~\AA~ waveband of the Atmospheric Imaging Assembly (AIA; \citealp{Lemen2012}) on board SDO and Extreme UltraViolet Imager (EUVI; \citealp{Wuelser2004}) on board STEREO. AIA images the solar corona with a pixel size of $0\arcsec.6$ and a high cadence of 12 s whereas EUVI images from STEREO have a pixel size of $1\arcsec.6$ and obtained at a cadence of 10 minutes. Using this 3D coordinate measuring tool, the precise 3D positions of the prominence feature is determined (see section~\ref{ch_di}).

Further the decay index, n(z) of the background field is computed by reconstructing the coronal magnetic field using potential field source surface model (PFSS; \citealp{Schrijver2003}). PFSS approximates the coronal magnetic field as potential field between the photosphere and spherical surface at $2.5R_{\odot}$ and the magnetic field on the spherical surface is radial. This model is implemented in PFSS package available in Solarsoftware (SSW). The model requires the radial magnetic field at the photosphere as boundary condition, which is the daily-updated synoptic chart of the photospheric radial magnetic field observations of the Helioseismic and Magnetic Imager (HMI; \citealp{Schou2012}) on board SDO. Each daily-updated synoptic chart is composed of two parts \citet{Sun2018}. The update part is a 120 degree wide band from a 4-hour average of the remapped magnetograms centered at the central meridian time of interest. The 120-degree updated region provides data in longitude from the left-edge towards the right. The remainder of the map comes from the standard Carrington synoptic chart(s) that makes up rest of the Carrington rotation.

\section{Analysis and Results}
\label{Res}

\subsection{Overview of sample events}

\begin{figure*}[!ht]
	\centering
	\includegraphics[width=.99\textwidth,clip=]{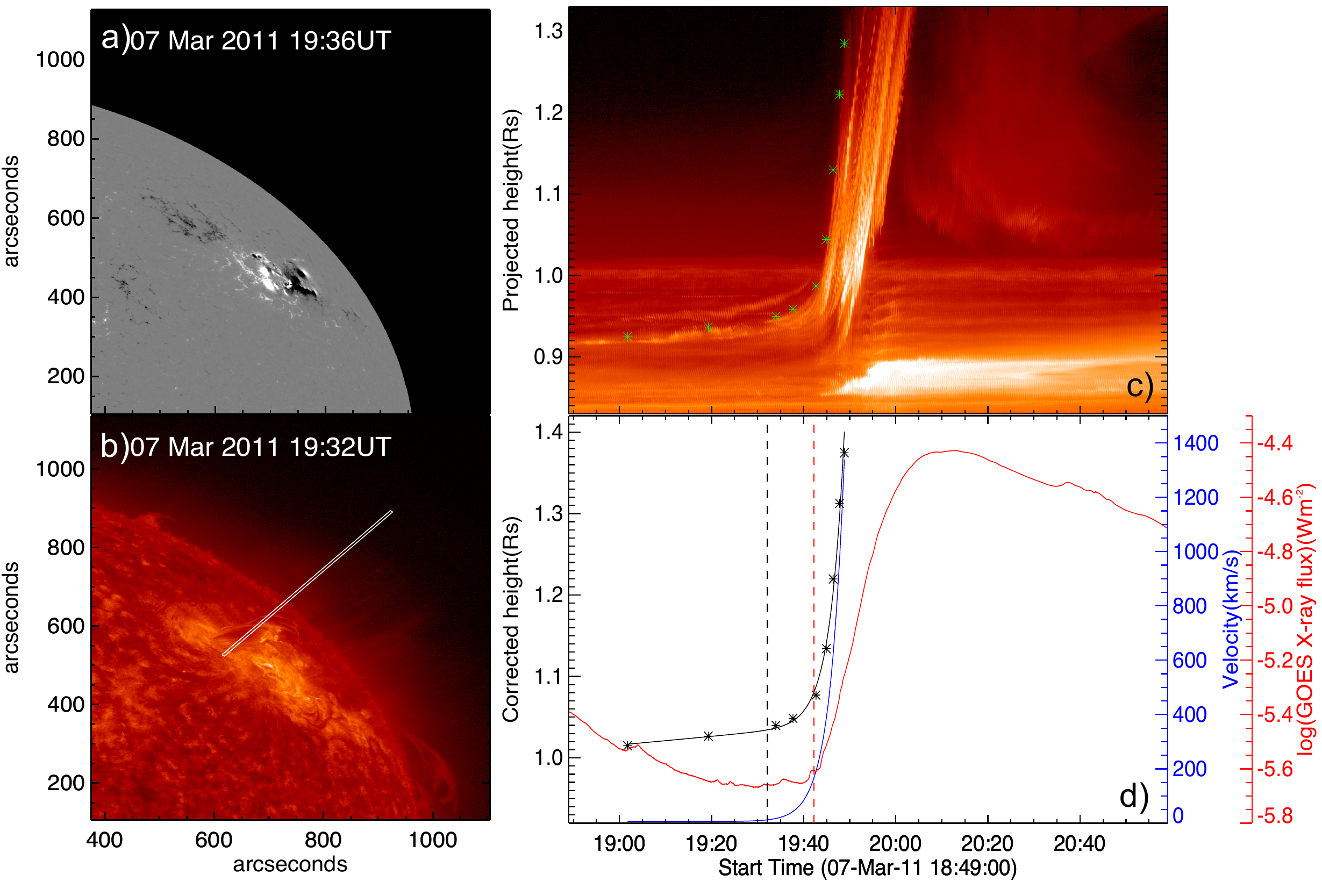}
	\caption{(a-b) HMI LOS magnetogram and AIA 304~\AA~ image of the prominence eruption event P1. A slit (in panel b) is placed to track the ascending prominence apex. (c) height-time stack image of the slit. Green asterisk symbols locate the data points for the ascending apex. (d) model fitting to the  corrected height-time data (black asterisks, black solid line). Blue solid curve is derived velocity. GOES soft X-ray flux is plotted in red solid curve. Red vertical dashed line (19:42 UT) marks the onset of the M3.7 flare associated with this eruption. Black vertical dashed line marks the time of onset (19:33 UT) of fast-rise phase.}
	\label{Fig_20110307}
\end{figure*}
The event P2 was erupted near the west limb above the neutral region between extended bipolar region adjacent to AR 11207 on 11 May, 2011 at 02:16~UT. This event is assiciated with two-ribbon flare B8.1 started to occur after the eruption i.e., at 02:23~UT and the eruption transitions into a partial halo CME. P4 event is a filament eruption from the AR 11343 in the north-eastern quadrant. During the rise motion, an apparent (un)twisting in the eastern footpoint of the filament is observed. Eventually the filament erupts at 12:57~UT on 9 November 2011, leading to a CME associated with a M-class flare. Event P5 is a fast erupting filament from the small AR 11386 near the Eastern limb. The eruption occurs at 08:16~UT on 24 November 2011 exciting high coronal oscillations. The CME is associated with C5.2 flare. The P6 filament erupted from the AR 11639 located near east limb on 26 December 2012 at 17:02~UT. This is a weak eruption and fall back of some material were observed after the eruption as seen in AIA 304~\AA. During the eruption apparent untwisting of the southern footpoint is observed and eventually lead to a minor CME. No associated flare were recorded by GOES  and also no flare ribbons were observed during/after the eruption. P7 filament erupted from AR 11667 near the west limb on 11 February 2013 at 18:48UT. It injects material into an open flux region. While some material as observed in AIA 304 channel escapes the AIA field-of-view leading to a CME, the remainder seems to be temporarily suspended in the corona before sun-ward descent. A GOES B5.8 X-ray flare was recorded from the AR after the P7 eruption. The event P8 is a large-scale filament eruption on the North-West disk from an extended bipolar region located north of the AR 11691. The ejected material floats around in the corona like a ``cloud'' for a long time after its eruption at 13:38~UT on 16 March 2013. In the beginning, the footpoints are bright enough to cause diffraction pattern in AIA 304~\AA, however there is no GOES X-ray flare associated with this event. The P9 is a bright and irregular-shaped prominence, erupted at 16:04~UT on 8 July 2014 from AR 12113 located near Eastern limb. The eruption leads to a fast partial halo CME associated with strong M6.5 flare. The P10 prominence appeared to have two branches of flux threads that are intermingled with each other towards the southern footpoint and the lower branch gets bifurcated at the apex to  a different footpoint during its rising motion. More detailed study of P10 prominence can be found in \citet{Vemareddy2017}. The P10 prominence erupted at 01:01~UT on 9  May 2015  from the AR 12342 located near East limb. The eruption manifests into a CME and associated with GOES C7.4 flare.
\begin{figure*}[!ht]
	\centering
	\includegraphics[width=.99\textwidth,clip=]{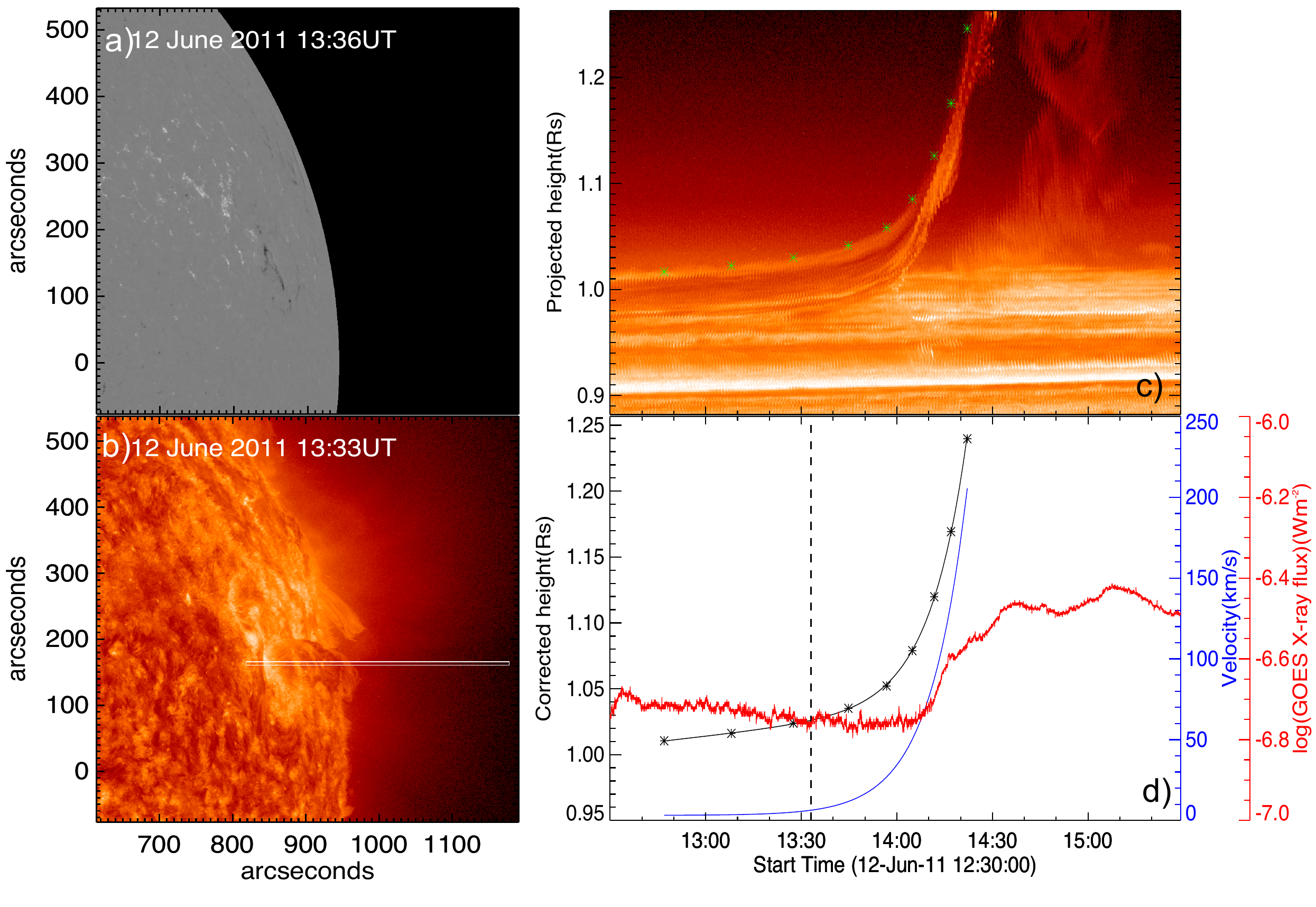}
	\caption{Panels (a-d): Same as figure~\ref{Fig_20110307} but for P3 event. P3 event is not associated with any GOES class flare.  Black vertical dashed line  marks the time of onset of  fast-rise phase i.e., at 13:34 UT.}
	\label{Fig_20110612}
\end{figure*}
 
 \subsection{Critical heights and Decay Indices}
 \label{ch_di}
As exemplary cases, we present events P1 (flare associated) and P3 (flare-less) to illustrate the procedure of determining the critical heights of erupting prominences and the corresponding decay indices.  The P1 erupted on Mar 7, 2011, at 19:33~UT from NOAA active region (AR) 11164 located near the west limb (N24 W62). This strong eruption happened before the M-class flare, which started at 19:42~UT and peaked at 20:12~UT according to GOES. Also during this time (19:42-20:58~UT) clear arcs are seen at the foot points of the flare in 304~\AA~image.  The P1 eruption appears to have a light bulb-shaped structure with the twisted filament in the middle and eventually leads to a halo CME observed in LASCO C2/C3 FOV. A detailed study of P1 is presented in \citet{Cheng2013a}. The P3 erupted on June 12, 2011, at 13:34 UT from AR NOAA 11232 located near the west limb (N08W74). The P3 eruption displayed a large kink associated with no observable X-ray flare as recorded by GOES and mass leaving the surface leads to CME observed in LASCO C2 field-of-view. The enhancement of SXR profile is observed only after 40 minutes of P3 eruption (Fig~\ref{Fig_20110612}d) and this small SXR intensity enhancement of B-class level is accounted for the brightening caused by some of the erupted mass when it falls back to the surface.

\begin{figure*}[!ht]
	\centering
	\includegraphics[width=.99\textwidth,clip=]{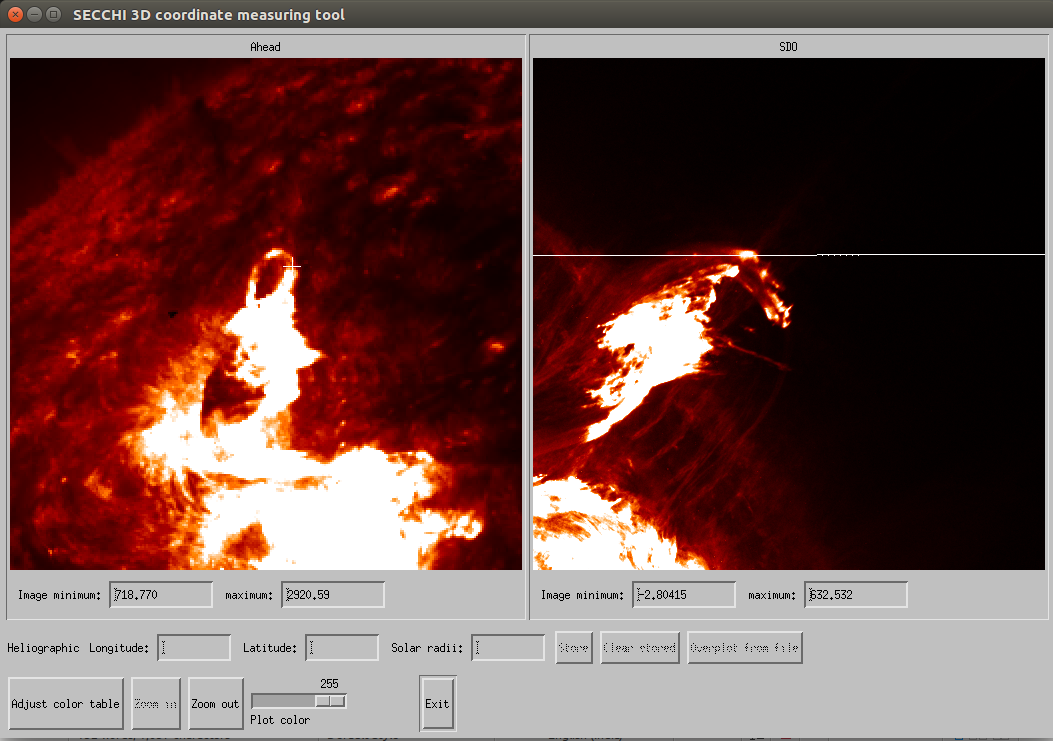}
	\caption{Screenshot of SECCHI 3D coordinate measuring interface. STEREO-A  and SDO  images are in left and right panels respectively during P1 eruption. The white horizontal line in SDO image is the epipolar line for the prominence feature position marked with ``+" sign on the STEREO-A image.}
	\label{Fig3}
\end{figure*}
Panels in Figure~\ref{Fig_20110307}(a~\&~b) and Figure~\ref{Fig_20110612}(a~\&~b) show the HMI Line-of-sight (LOS) magnetic field and AIA 304~\AA~observations of the prominence eruptions of P1 and P3 respectively. The slits were placed on the respective AIA 304~\AA~images to characterize the overall trajectory of erupting prominences. Using these slits, space-time plots were generated (Fig.~\ref{Fig_20110307}c \& Fig.~\ref{Fig_20110612}c) and green asterisk symbols in these panels represent the leading edge of ascending prominences. We applied the height correction using the 3D coordinates obtained from tie-pointing method. The SDO, STEREO-A and -B spacecrafts positions during P1 event were obtained using STEREO science center website (\url{https://stereo-ssc.nascom.nasa.gov/cgi-bin/make_where_gif}). We used near simultaneous observations of SDO/AIA  and STEREO-A  in wavelength passband of 304~\AA~in \texttt{scc\_measure.pro} routine and then with the aid of graphical interface of SECCHI 3D coordinate tool, we are allowed to select a feature on one image (see left panel of Fig.~\ref{Fig3}, ``+" sign on STEREO-A image), then an epipolar line will display on the SDO image passing through the same feature as shown in right panel of Figure~\ref{Fig3}. After the user identifies the intersection between the projected line of sight and the feature of interest, the program triangulates the feature's three-dimensional (3D) location. Using this 3D coordinate of the selected feature, the correction is added/subtracted to the projected height.

\begin{figure*}[!ht]
	\centering
	\includegraphics[width=.99\textwidth,clip=]{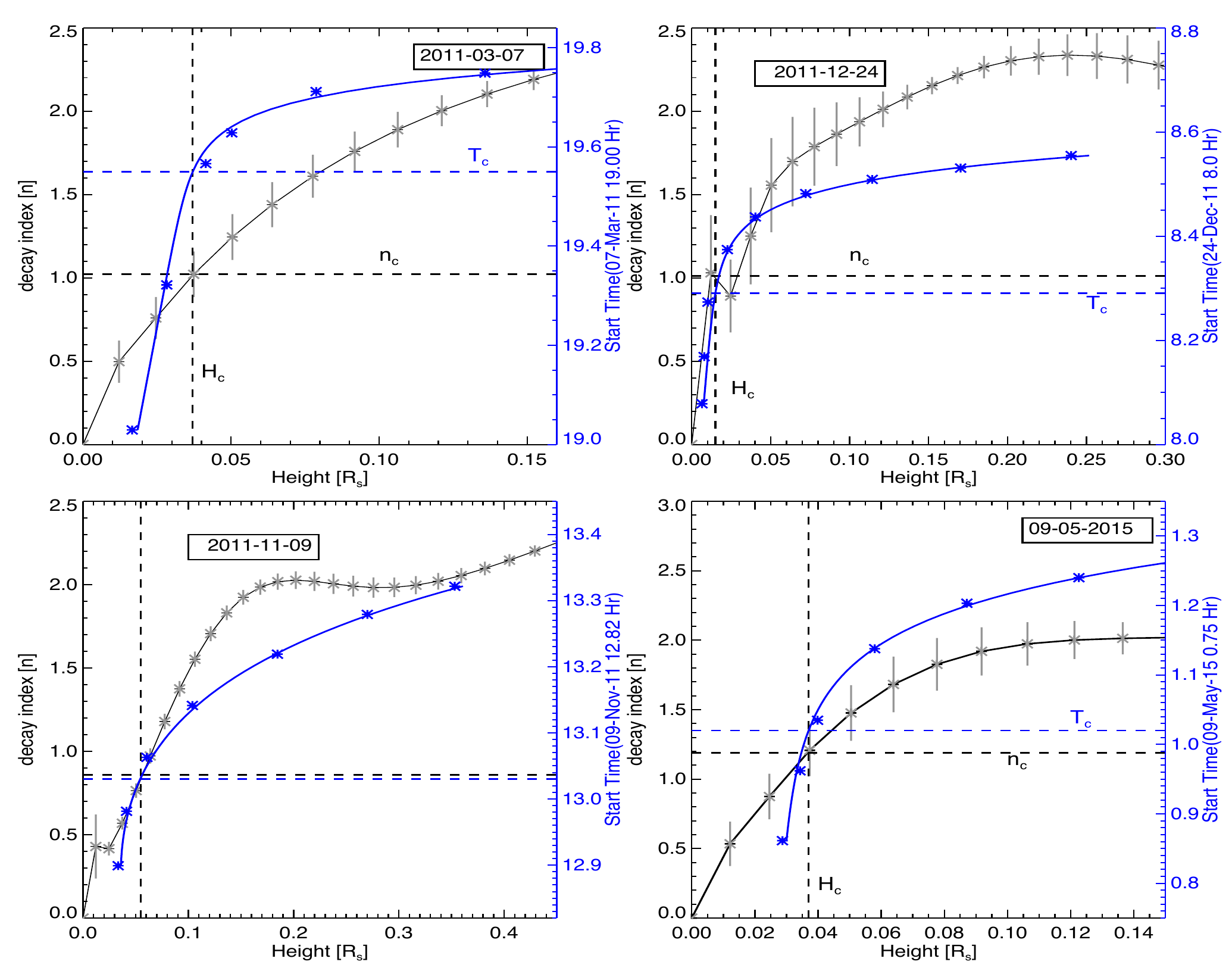}
	\caption{Variation of decay index with height above the photosphere during the P1 (top left), P5 (top right), P4 (bottom left), p10 (bottom right) events. Height-time profile is also plotted (blue curve) with time as y-axis scale. From the model fit, the critical height (vertical dashed line) is determined as the height at which erupting prominence commences the onset of the fast-rise motion. Corresponding to the critical height ($H_{c}=0.035R_{\odot}$) and time ($T_{c}=19.55Hr$), the $n_{c} = 1.02\pm0.12$.} 
	\label{Fig4}
\end{figure*}

\begin{figure*}[!ht]
	\centering
	\includegraphics[width=.99\textwidth,clip=]{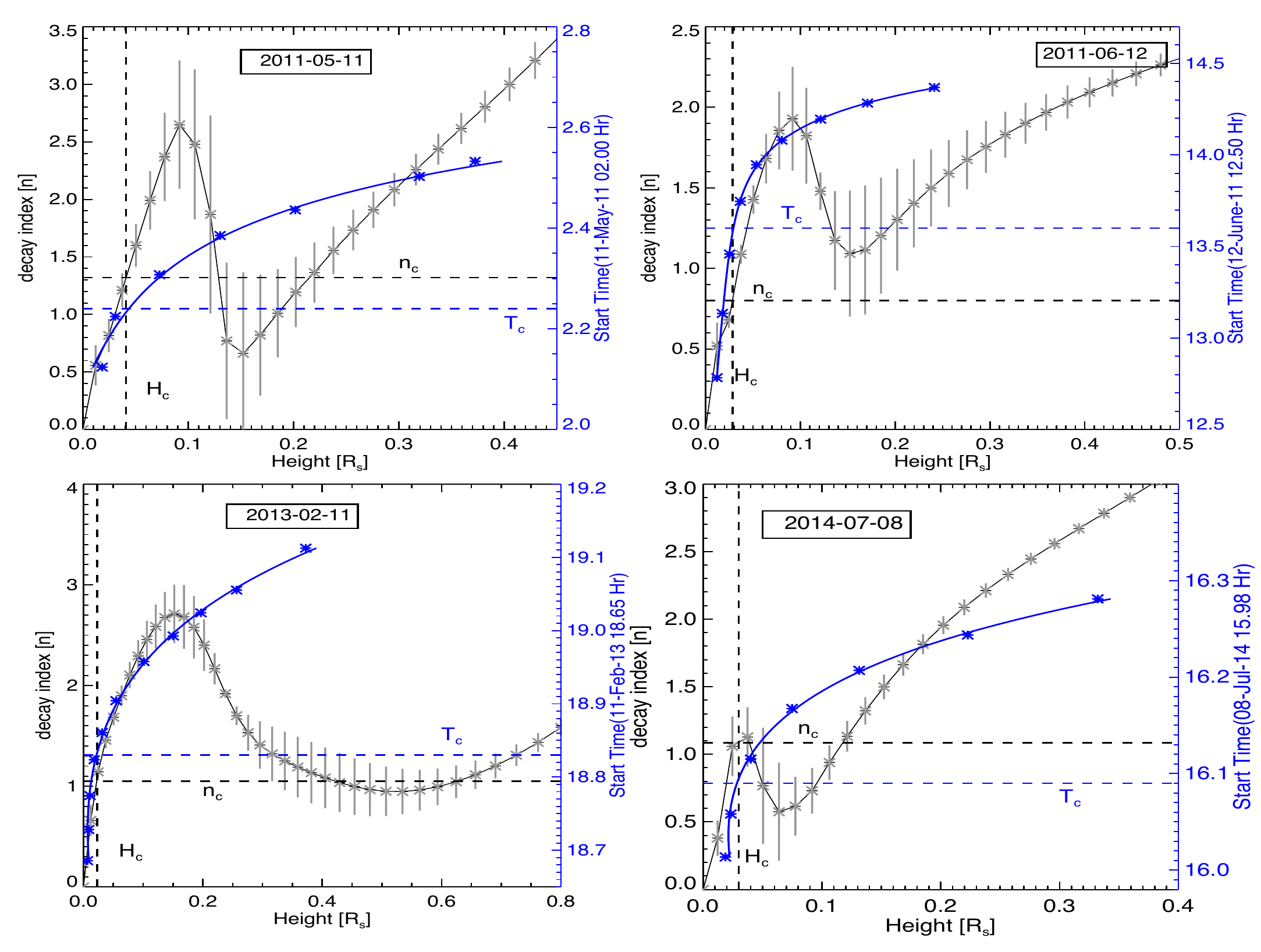}
	\caption{Same as figure~\ref{Fig4} but for different set of events: P2 (top left), P3 (top right), P7 (bottom left) and P9 (bottom right). Note that the variation of decay index is different than the set of events presented in figure~\ref{Fig4}. A bump in the decay index curve is observed in low corona for these set of events.} 
	\label{Fig5}
\end{figure*}

The de-projected or corrected height-time plots of P1 and P3 are shown in the Figures~\ref{Fig_20110307}(d) and ~\ref{Fig_20110612}(d) respectively. The corrected height-time curve consists of two distinct profiles, a slow-rise phase having almost constant velocity (e.g. \citealp{Sterling2005}) and a fast-rise phase with rapid acceleration approximated by exponential curve (e.g., \citealp{Goff2005}). We used a model containing the linear term to treat the slow-rise phase and exponential term to account for rapid-acceleration phase as described in \citet{Cheng2013b} and is given by 
\begin{equation}
 h(t) = C_{0} e^{(t-t_{0} )/\tau} + C_{1} (t-t_{0} ) + C_{2}   
\end{equation}
where $h(t)$ is height at a given time t, and $\tau$, $t_{0}$, $C_{0}$, $C_{1}$, and $C_{2}$ are free coefficients. This model has two distinct advantages: (1) a single function describes the two phases of eruption effectively and (2) it provides a convenient method to determine the time of onset of rapid-acceleration phase ($T_{c}$). Critical time, $T_{c}$ is defined as the time at which the exponential component of velocity equals to its  linear component as $T_{c} = \tau ln(C_{1}\tau /C_{0} ) + t_{0}$. Using this equation, $T_{c}$ for P1 and P3 events are computed to be 19:33 UT (19.55Hr) and 13:34 UT (13.57Hr) respectively. These timings are marked by blue horizontal dashed lines in Figures~\ref{Fig4}a and ~\ref{Fig5}b. We used \texttt{mpfit.pro} to fit the corrected height-time data by model function and the fit is showed as blue solid curve. From this fit, the critical height ($H_{c}$) is determined corresponding to $T_{c}$. The $H_{c}$ for P1 and P3 events are determined to be $0.035R_{\odot}$ \& $0.028R_{\odot}$ respectively and they are represented by black vertical dashed lines in Figures~\ref{Fig4}a and ~\ref{Fig5}b.

Further to determine the decay index corresponding to the critical height, we used HMI daily-updated synoptic maps as the boundary conditions in potential field approximation (PFSS). After extrapolating, horizontal component of background magnetic field ($B_{h}$) as a function of height is obtained at eight to ten points along the main polarity inversion line (PIL). Then an average of {\em n} is derived. Errors of {\em n} are mainly from the uncertainties in height, which are regarded as the standard deviations of number of measurements. Then using the decay index curve, decay index corresponding to critical height ($H_{c}$) is determined and considered as $n_{c}$.  The critical decay indices of P1 and P3 events are found to be 1.02 \& 0.8 respectively and are represented by black horizontal dashed lines in Figures~\ref{Fig4}a and ~\ref{Fig5}b. We followed the same procedure to derive the parameters like $T_{c}$, $H_{c}$ and $n_{c}$ for all the ten events in our sample, which are tabulated in Table~\ref{tab1}. The average critical decay indices of our sample of ten events is $\approx1.05$. This result is in consistent with the past studies like \citet{Filippov2001} and \citet{McCauley2015} (see introduction). But both these studies involve the errors induced by the projection effects on the determination of prominence positions. To account for these errors, \citet{Filippov2013ApJ} used three vantage point observations to study the quiescent filament eruption and found that $n_{c}$ is $\approx1.0$. Thus, generally, $n_{c}$ for both quiescent and AR prominences are almost equal to 1 and its value prominently depends on the strength of coronal background magnetic field confinement of individual event.

We observed two types of decay index curve in our sample events. Four (P1, P4, P5 \& P10) out of ten events exhibit gradual increase of decay index with height and are shown in Figure~\ref{Fig4}. For remaining six events, the decay index curves exhibit the unusual `bump' in low corona and then increases gradually with height. These events are shown in Figure~\ref{Fig5}. The critical heights of our sample events are less than 40 Mm (height above the photosphere) except for P8 event, which has the critical height at about 50 Mm. The obtained critical heights are in agreement with recent study of \citet{Vasantharaju2018} where majority of critical heights of eruptive events are less than 42 Mm. Corresponding to the critical height ($H_c$), the $n_{c}$ of our sample events are in the range of 0.8 to 1.3.

Thus the background magnetic field in the lower corona ($< 50 Mm$) decreases quite fast enough over all the eruption sites of our sample events to facilitate the eruptions irrespective of type of decay index profiles. However, in the events where the decay index curves exhibit bump in the lower corona, the background transverse magnetic field decreases very rapidly than over the sites where decay index curves show gradual increase. For example the events like P2, P3 and P7 events (except P9) exhibit bump in decay index curves in lower corona (i.e.,within 50 Mm), the maximum decay index values reached are in the range of 1.5 to 2.0 (Fig.~\ref{Fig5}) but for events like P1, P4 ,P5 \& P10 which do not show bump in decay index curves, the maximum decay index values attained are smaller and in the range of 1.1 to 1.6 (Fig.~\ref{Fig4}). Though P9 event exhibits a bump in the decay index curve (Figure~\ref{Fig5}d) in lower corona, the maximum decay index value attained is just about 1.2 at the height of about 20 Mm. Also, the P9 eruption got initiated at about 20 Mm (blue curve in Figure~\ref{Fig5}d). This strongly suggests that the enough weaker transverse magnetic field strength at about 20 Mm facilitates the eruption or initiates the fast-rise motion of erupting structure. After its eruption the decay index value decreases and then gradually increases with height. The decay index variations with height for P6, P8 (both figures are not shown) and P7 (Figure~\ref{Fig5}c) events are almost similar and they exhibit a large bump in the decay index curves representing large decay index values (1.5 - 2.5) upto the height of more than 120 Mm. This very rapid decay of background field in the lower corona initiates the eruptions of flux ropes/erupting structures in these events. Then the decay index values decreases slowly with height and again it starts to rise after about 300 Mm.

\subsection{Kinematics of prominences}

\begin{figure*}[!ht]
	\centering
	\includegraphics[width=.99\textwidth,clip=]{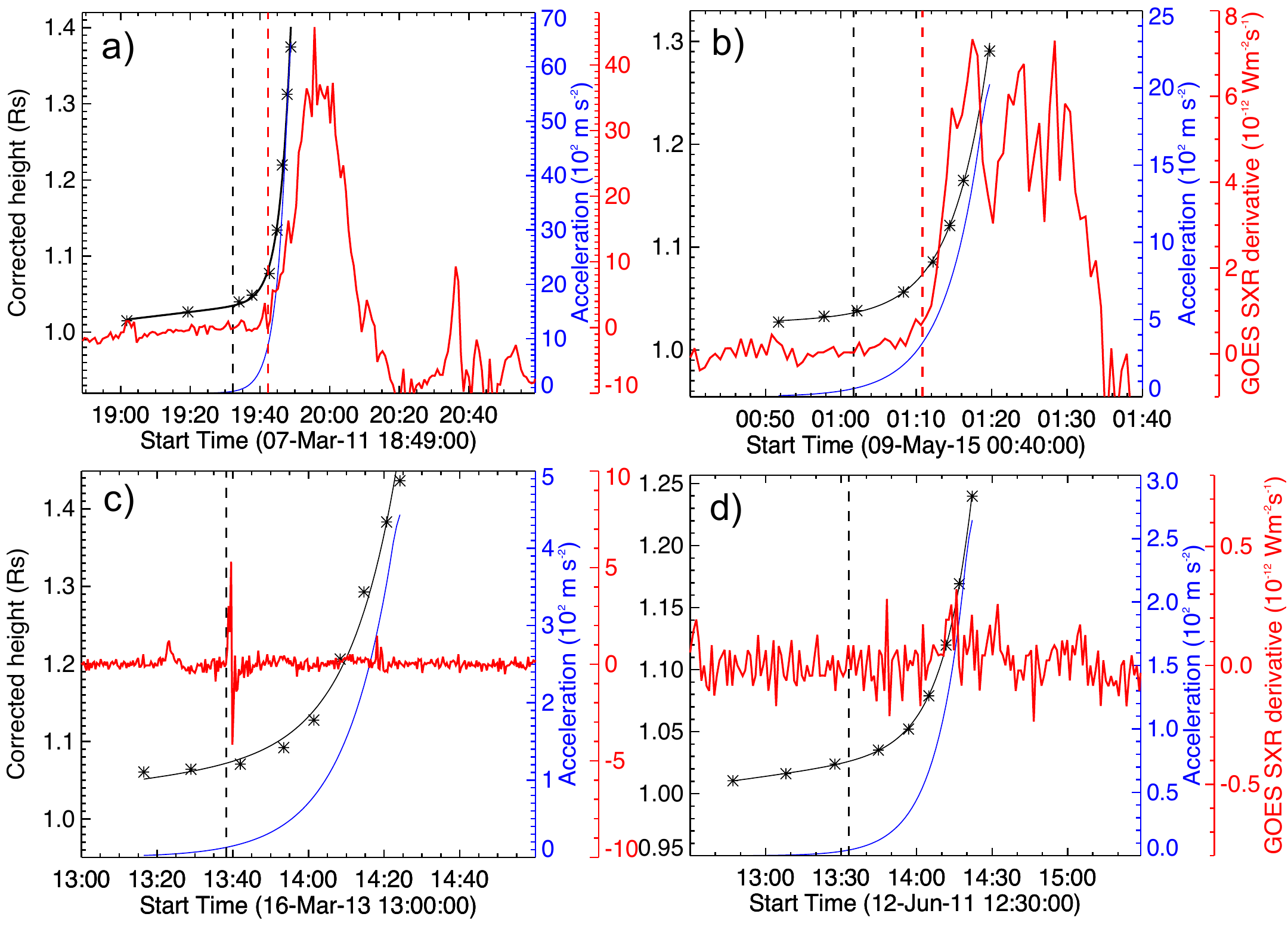}
	\caption{(a): Comparison of the height-time profile of the prominence eruption with the GOES soft X-ray flux. a) P1 event, b) P10 event, c) P8 event, d) P3 event. The derived acceleration (blue curve) and GOES X-ray flux \textbf{derivative} (red curve) are shown in the same panels with y-axis scale on right. Black vertical dashed line marks the time of onset of fast-rise phase. Red vertical dashed line indicates the time of flare onset. Top two panels are flare-associated events and bottom two panels are for flare-less events. Clearly, flare-less events have smaller acceleration typically by a factor of ten compared to the flare-associated events.} 
	\label{Fig6}
\end{figure*}

All our sample events exhibit distinct slow- and fast- rise phases which are well-fitted by the linear-cum-exponential model proposed by \citet{Cheng2013b}. The two representative events P1 and P3 used in previous section will also be considered here along with P10 and P8 to illustrate the kinematics of our sample events. P1 and P10 are flare accompanied prominence eruptions whereas P3 and P8 are flare-less events. The velocity and acceleration profiles are obtained by taking time derivatives of model function fit to the corrected height-time data. In Figure~\ref{Fig6}, top panels present the typical examples (P1 \& P10) of kinematics of flare-associated events and bottom panels present the kinematics of flare-less prominence eruption events (P3 \& P8). Solid black curve indicates the fit to corrected height-time data points (black asterisks in fig.~\ref{Fig6}) and the derived acceleration profiles are over-plotted in blue in all four panels. The critical time obtained from the fit bifurcates the prominence rise motion into slow- and fast-rise (rapid-acceleration) phases. Thus, the slow-rise phase for P1 and P10 is observed till 19:33~UT on 7 March, 2011 and 01:01~UT on 9 May, \textbf{2015} with average velocities of ~7 km/s and ~9 km/s respectively. For P3 and P8 events, the slow-rise phase till 13:34~UT on 12 June, 2011 and 13:38~UT on 16 March, 2013 with the average velocities of ~4 km/s and ~9 km/s respectively. After that, fast-rise phase starts for flare-associated events P1 and P10 with average acceleration of~1541~$ms^{-2}$ and~578~$ms^{-2}$ respectively. For flare-less events P3 and P8, the average acceleration during rapid-acceleration phase is~72~$ms^{-2}$ and~128~$ms^{-2}$ respectively. Similarly, kinematic details of other events are tabulated in Table~\ref{tab2}. In our sample, we found that in rapid-acceleration phase, the flare associated events have average acceleration in the range~400~$ms^{-2}$ to 1550~$ms^{-2}$ and flare-less events (P3, P6 \& P8) have average acceleration well below 200~$ms^{-2}$. These observational results imply that the flare magnetic-reconnection occurred during fast-rise phase is responsible for the acceleration of flare-associated prominences significantly compared to that of flare-less counterparts.

Further, the GOES soft X-ray derivative can be used as proxy of hard X-ray according to the Neupert effect \citep{Neupert1968} and is over plotted in red in all panels of  Figure~\ref{Fig6} to identify the time of onset of magnetic reconnection (flare). The time of onset of flare during P1 and P10 events are 19:42~UT and 01:11 UT respectively, which are about 9-10 minutes later than the time of onset of fast-rise phases (i.e., 19:33 UT and 01:01UT) of P1 and P10 respectively. It is worth to note that in all our flare-associated events, onset time of fast-rise phase is earlier than the onset time of flare by 2-10 minutes (see  Table~\ref{tab2}). After careful inspection of temporal correlations between the velocity and soft X-ray flux as well as the acceleration and hard X-ray flux, we noticed that the flux-rope instability triggers and accelerates the prominences first, and the magnetic reconnection is induced subsequently to provide further acceleration as suggested in past studies \citep{Priest2002,Lin2003,Temmer2010,Vrsnak2016}. 
\begin{table*}
	\caption{Kinematics of prominence eruption events. $T_c$: Onset time of fast-rise phase;
		$T_f$: Onset time of flare accompanied;
		$V_{avg}$: Average velocity of slow-rise phase;
		$V_{max}$: Maximum velocity in field-of-view(FOV) of slice; 
		$a_{ini}$: Acceleration at the onset ($T_c$)  of fast-rise phase;
		$a_{max}$: Maximum acceleration in FOV  of slice;
		$a_{avg}$: Average acceleration of fast-rise phase; 
		$a_{pf}$:Average acceleration of pre-flare phase;
		$a_{ip}$:Average acceleration of impulsive phase.}
	\begin{center}
		\small\addtolength{\tabcolsep}{-1.5pt}
		\begin{tabular}{|c|c|c|c|c|c|c|c|c|c|}
			\hline 
			P No.    &    $T_c$       &       $T_f$      &     $V_{avg} (km/s)$     &    $V_{max} (km/s)$     &     $a_{ini} (ms^{-2})$      &    $a_{max} (ms^{-2})$     &      $a_{avg} (ms^{-2})$       &      $a_{pf} (ms^{-2})$       &     $a_{ip} (ms^{-2})$ \\ 
			\hline 
			P1     &     19:33  &       19:42  &      7.1 &       1347.1  &     30.1  &     6441.2 &    1540.8   &    299.2  &   2970 .3 \\
			P2    &      02:16   &      02:23  &     30.2 &       713.5  &     109.2  &    1273.1 &      532.6   &    235.8   &     726.3 \\
			P3      &    13:33   &         --  &          3.9   &      205.6   &      4.5    &     264.6 &        71.6      &    --    &          --\\
			P4     &     12:57    &     13:04   &    18.8     &    670.8  &     54.6  &      911.4  &      397.1    &     90.3    &    499.8\\
			P5     &     08:16   &      08:27  &     10.8   &      617.3   &    37.8   &    2389.1  &     649.7   &    201.6  &    1301.3\\
			P6      &    17:02     &       --  &         30.3     &    315.3   &    67.7   &     616.5 &      198.6   &       --    &           --\\
			P7     &     18:48    &     18:55   &    24.4   &      696.7  &     90.6   &    1676.3 &      591.2  &     191.6   &     808.9\\
			P8      &    13:38   &          --&            8.6  &       359.3 &      13.1   &     422.8&       128.2    &       --  &           --\\
			P9      &    16:04  &       16:06 &       52.4    &     982.6    &  220.4  &   1683.2 &      839.7  &      247.4   &     928.6  \\
			P10      &   01:01 &        01:11 &        8.9    &     635.6   &    47.2  &    2023.3 &      577.2    &    166.7  &    1020.3\\   
			\hline 
		\end{tabular} 
	\end{center}
	\label{tab2}
\end{table*}

Further to compare the mechanisms contributing to the acceleration process during the fast-rise phase of flare-associated events (impulsive events) in AIA FOV, the fast-rise phase can be divided into two phases viz. a pre-flare and flare-impulsive phases, based on the reference time of onset of flare during the events. Here, the pre-flare phase is assumed as the duration of the fast-rise phase from the critical time of eruption to the time of onset of flare and rest of the fast-rise phase i.e., from the time of onset of flare till the prominence structure leaves the AIA FOV is considered as an impulsive phase. Though actually the pre-flare phase is combination of slow-rise phase and the fast-rise phase till the flare onset, we have excluded the slow-rise phase due to the fact that the acceleration process in slow-rise phase is contributed by both kink-instability and/or quasi-separatrix-layer (QSL) reconnection \citep{Cheng2013b}, i.e., reconnection at the interface between the filament and its surrounding corona. As it is difficult to disentangle the contributory mechanisms to acceleration process in slow-rise phase, we deliberately excluded it in considering from pre-flare phase. So basically, we are concentrating on the mechanisms contributing to the acceleration process from the prominence's eruption to till it leaves the AIA FOV. In figure~\ref{Fig6}, the enhancement of the SXR flux derivative as indicated by the red vertical dashed lines mark the onset time of flare magnetic-reconnection in top panels and steady behaviour of SXR flux derivative in the bottom panels indicate the absence of flare during prominence eruptions. The enhancement of SXR flux derivative for P1 (Fig.~\ref{Fig6}a) and P10 (Fig.~\ref{Fig6}b) events occurred at 19:42 UT and 01:11~UT respectively indicate the time of onset of impulsive phase. The average acceleration in the pre-flare phase of P1 event is about 299~$ms^{-2}$ and in impulsive phase is about  2970~$ms^{-2}$. For P10 event, the average acceleration during pre-flare phase and impulsive phase is ~$167 ms^{-2}$ and ~$1020ms^{-2}$ respectively. The average acceleration in impulsive phase is almost 10 times larger than that of pre-flare phase for P1 event and in the same way for P10 event its almost 6 times greater. Similar to P1 and P10 events, remaining flare-associated events are highly accelerated in their impulsive phases than in pre-flare phases and their kinematic parameters are listed in Table~\ref{tab2}. These results suggests that magnetic reconnection is the major contributor to the acceleration process than the torus instability in impulsive phases of prominence eruptions. Also, by considering the whole acceleration process during fast-rise phase, the contribution of magnetic reconnection to the acceleration of flux-rope is dominant than the flux-rope instability and not these two mechanisms have comparable contributions to the acceleration process in impulsive events. This result is in agreement with the recent analytical study of \citet{Vrsnak2016}.

\section{Summary and Discussion}  
\label{summ}
The background field is assumed to play prime role in the erupting structures like prominences. In the flux rope models, the $n_{c}$ is a measure of vertical gradient of the background magnetic field and an important dimensionless parameter determining the eruptive nature of prominences. Owing to difficulties in obtaining the critical height of the background field, a typical value of $n_{c}=1.5$ is adopted from the numerical studies (eg. \citealp{Torok2005}). In this study, we investigated the critical decay indices of 10 prominence eruption events, by estimating the critical height from two vantage-point observations of the erupting prominence. 

Ideal MHD instabilities responsible for the flux rope eruption include the torus instability and helical kink instability. For a flux rope of exceeding magnetic twist, the kink-instability may drive the eruption upto a point of torus regime. Kink-instability is not necessarily be a trigger in all events, it can be tether-cutting reconnection. When the flux rope reached to a height of steep gradient of horizontal field strength, both the instabilities may contribute to the impulsive acceleration of the prominence simultaneously, even along with the reconnection. At a height of critical point, the downward force is dominated by the hoop force, leading to the commencement of flux rope fast rise motion. Therefore, we argue that the critical height of steep field strength gradient is the height of onset of fast-rise motion of the prominence.

Our sample events exhibit a linear slow- and exponential fast-rise phases during the eruption and the decay indices are determined at the onset height of fast-rise phase. The assumption, as justified earlier, involved is that at the time the prominence commences the fast-rise motion, it (the apex) is at a critical height of background field. From the critical height, the erupting structure is allowed to expand exponentially provided there is no strapping background field. In this scenario, the height-time profile of the erupting structure exhibits a turning point from slow to fast rise motion. We use a fitting model to determine this turning point as critical height. The background field is obtained from PFSS model by using HMI synoptic magnetic map. 

Two types of decay index are observed in our 10 sample events. Four events exhibit the gradual increase of decay index with height and remaining six events exhibit a `bump' in decay index curves in lower corona. This unusual bump in decay index curves were first reported in \citet{Cheng2011} and claimed to be mostly seen in the eruptive flare cases. However, we observed similar bump for flare-less prominence eruptions (P6 \& P8) as well. These bumps in the lower corona represent a weaker transverse magnetic field or very rapid decrease of background magnetic field facilitating the eruptions (initiating the fast-rise motion of eruptive structure) at critical heights. 

The $n_{c}$ is not a constant value and varies from event to event as the background field configuration depends on the field distribution in the source region. The $n_{c}$ of our sample events ranges from 0.8 -- 1.3 and the average value of $n_{c}$ of our sample is found to be 1.05. This value is in agreement with the past observational studies \citep{McCauley2015,Filippov2013ApJ}. Any differences in critical decay indices obtained from theoretical and observational studies is just an apparent and this is mainly due to the location where exactly the torus instability criterion is evaluated. Numerical simulations by \citet{Zuccarello2016} showed that in curved flux rope geometry, the height of the flux rope axis is larger than the apex of prominence structure. Due to this, the $n_{c}$ obtained at the height of flux rope axis in simulations and the apparent $n_{c}$ obtained at the top of dipped structure using observations are found to be $1.4\pm0.1$ and $1.1\pm0.1$ respectively.

The critical times are the onset time of fast-rise (rapid-acceleration) phases, which are obtained using linear-exponential model of \citet{Cheng2013b}. In all our sample events, the onset time of flares is 2-10 minutes later than the onset time of rapid-acceleration phases. We observed that prominence events associated with flares are highly accelerated compared to flare-less counterparts. 

To study the comparison of mechanisms contributing to the acceleration process within AIA FOV of flare-associated (impulsive) events, the fast-rise phase is further separated into two sub-phases based on the onset time of the flares viz. a pre-flare phase (excluding slow rise motion) and an impulsive phase. We inferred from temporal correlations between the velocity and soft X-ray flux, as well as the acceleration and hard X-ray flux, the flux-rope instability holds a major contribution to the initial phase of acceleration in pre-flare phase and the flare magnetic-reconnection was dominant in the second phase i.e, impulsive phase. This analysis further leads to infer that the magnetic reconnection is the dominant contributor to acceleration process than the MHD instability within AIA FOV of impulsively accelerated prominence events. The weak magnetic reconnection in the pre-flare phase might be insufficient to accelerate the MFR as it cannot weaken the tension force of the overlying magnetic loops fast enough, and may even lead to the MFR deceleration as calculated by \citet{Lin2000}. However, the increase in the GOES SXR flux derivative (proxy of hard X-rays) indicates the increased flare reconnection rate that may lead to rapid decrease in the magnetic tension of overlying loops which in turn leads to the enhanced outward acceleration of prominence structure. Also, the analytical study of \citet{Vrsnak2016} shows that magnetic reconnection not only reduces the tension of overlying magnetic loops and increases the magnetic pressure below the ejecting flux rope but also supplies the additional poloidal flux to the flux rope and increases its hoop force. These factors enhance and prolong the flux rope acceleration significantly. Whereas for three flare-less events (P3, P6 \& P8) in our sample, the ideal MHD instabilities appear to be the dominant contributor to the acceleration process of erupting prominence structures. Numerical study by \citet{Chen2007a} showed that the ideal MHD instability process alone can produce fast CMEs but not faster than impulsive events. They further showed that if the magnetic reconnection sets in then it enhances the CME acceleration significantly, and both ideal MHD instability and magnetic reconnection have comparable contributions to the acceleration process in impulsive events. This notion is supported by observational studies of filament eruptions from both the active region \citep{Song2015} and the quiet region \citep{Song2018a}. However, the scenario is different from event to event. For example, in a statistical study of CME kinematics of 22 events \citep{Maricic2007}, a quarter of sample events exhibits the weak-synchronization of CME kinematics and magnetic-reconnection characteristics. In such events, the ideal MHD instability would be the major contributor to the CME acceleration.


\acknowledgements SDO is a mission of NASA's Living With a Star Program. Full-disk EUVI images are supplied courtesy of the STEREO Sun Earth Connection Coronal and Heliospheric Investigation (SECCHI) team. N.V is a CSIR-SRF, gratefully acknowledges the funding from CSIR-HRDG, New Delhi. We thank the referee for his/her encouraging comments and suggestions.


\begin{thebibliography}{56}
	\expandafter\ifx\csname natexlab\endcsname\relax\def\natexlab#1{#1}\fi
	
	\bibitem[{{Aulanier} \& {Demoulin}(1998)}]{Aulanier1998}
	{Aulanier}, G., \& {Demoulin}, P. 1998, \aap, 329, 1125
	
	\bibitem[{{Bateman}(1978)}]{Bateman1978}
	{Bateman}, G. 1978, {MHD instabilities} (Cambridge, Mass., MIT Press, 1978.~270
	p.)
	
	\bibitem[{{Chen}(2011)}]{Chen2011}
	{Chen}, P.~F. 2011, Living Reviews in Solar Physics, 8, 1
	
	\bibitem[{{Chen} {et~al.}(2007{\natexlab{a}}){Chen}, {Hu}, \&
		{Sun}}]{Chen2007a}
	{Chen}, Y., {Hu}, Y.~Q., \& {Sun}, S.~J. 2007{\natexlab{a}}, \apj, 665, 1421
	
	\bibitem[{{Chen} {et~al.}(2007{\natexlab{b}}){Chen}, {Hu}, \&
		{Xia}}]{Chen2007b}
	{Chen}, Y., {Hu}, Y.~Q., \& {Xia}, L.~D. 2007{\natexlab{b}}, Advances in Space
	Research, 40, 1780
	
	\bibitem[{{Cheng} {et~al.}(2013{\natexlab{a}}){Cheng}, {Zhang}, {Ding}, \& {et
			al}}]{Cheng2013a}
	{Cheng}, X., {Zhang}, J., {Ding}, M.~D., \& {et al}. 2013{\natexlab{a}}, \apj,
	763, 43
	
	\bibitem[{{Cheng} {et~al.}(2011){Cheng}, {Zhang}, {Ding}, {Guo}, \&
		{Su}}]{Cheng2011}
	{Cheng}, X., {Zhang}, J., {Ding}, M.~D., {Guo}, Y., \& {Su}, J.~T. 2011, \apj,
	732, 87
	
	\bibitem[{{Cheng} {et~al.}(2013{\natexlab{b}}){Cheng}, {Zhang}, {Ding},
		{Olmedo}, {Sun}, {Guo}, \& {Liu}}]{Cheng2013b}
	{Cheng}, X., {Zhang}, J., {Ding}, M.~D., {Olmedo}, O., {Sun}, X.~D., {Guo}, Y.,
	\& {Liu}, Y. 2013{\natexlab{b}}, \apj, 769, L25
	
	\bibitem[{{D{\'e}moulin} \& {Aulanier}(2010)}]{Demoulin2010}
	{D{\'e}moulin}, P., \& {Aulanier}, G. 2010, \apj, 718, 1388
	
	\bibitem[{{Fan}(2010)}]{Fan2010}
	{Fan}, Y. 2010, \apj, 719, 728
	
	\bibitem[{{Fan} \& {Gibson}(2007)}]{Fan2007}
	{Fan}, Y., \& {Gibson}, S.~E. 2007, \apj, 668, 1232
	
	\bibitem[{{Filippov}(2013)}]{Filippov2013ApJ}
	{Filippov}, B. 2013, \apj, 773, 10
	
	\bibitem[{{Filippov} \& {Den}(2001)}]{Filippov2001}
	{Filippov}, B.~P., \& {Den}, O.~G. 2001, \jgr, 106, 25177
	
	\bibitem[{{Goff} {et~al.}(2005){Goff}, {van Driel-Gesztelyi}, {Harra},
		{Matthews}, \& {Mandrini}}]{Goff2005}
	{Goff}, C.~P., {van Driel-Gesztelyi}, L., {Harra}, L.~K., {Matthews}, S.~A., \&
	{Mandrini}, C.~H. 2005, \aap, 434, 761
	
	\bibitem[{{Gosling}(1993)}]{Gosling1993}
	{Gosling}, J.~T. 1993, Physics of Fluids B, 5, 2638
	
	\bibitem[{{Gosling} {et~al.}(1974){Gosling}, {Hildner}, {MacQueen}, {Munro},
		{Poland}, \& {Ross}}]{Gosling1974}
	{Gosling}, J.~T., {Hildner}, E., {MacQueen}, R.~M., {Munro}, R.~H., {Poland},
	A.~I., \& {Ross}, C.~L. 1974, \jgr, 79, 4581
	
	\bibitem[{{Ji} {et~al.}(2003){Ji}, {Wang}, {Schmahl}, {Moon}, \&
		{Jiang}}]{Ji2003}
	{Ji}, H., {Wang}, H., {Schmahl}, E.~J., {Moon}, Y.~J., \& {Jiang}, Y. 2003,
	\apj, 595, L135
	
	\bibitem[{{Jing} {et~al.}(2018){Jing}, {Liu}, {Lee}, {Ji}, {Liu}, {Xu}, \&
		{Wang}}]{Jing2018}
	{Jing}, J., {Liu}, C., {Lee}, J., {Ji}, H., {Liu}, N., {Xu}, Y., \& {Wang}, H.
	2018, \apj, 864, 138
	
	\bibitem[{{Kahler}(1992)}]{Kahler1992}
	{Kahler}, S.~W. 1992, \araa, 30, 113
	
	\bibitem[{{Kippenhahn} \& {Schl{\"u}ter}(1957)}]{Kippenhahn1957}
	{Kippenhahn}, R., \& {Schl{\"u}ter}, A. 1957, \zap, 43, 36
	
	\bibitem[{{Kliem} \& {T{\"o}r{\"o}k}(2006)}]{Kliem2006}
	{Kliem}, B., \& {T{\"o}r{\"o}k}, T. 2006, Physical Review Letters, 96, 255002
	
	\bibitem[{{Lemen} {et~al.}(2012){Lemen}, {Title}, {Akin}, \& {et
			al}}]{Lemen2012}
	{Lemen}, J.~R., {Title}, A.~M., {Akin}, D.~J., \& {et al}. 2012, \solphys, 275,
	17
	
	\bibitem[{{Lin} \& {Forbes}(2000)}]{Lin2000}
	{Lin}, J., \& {Forbes}, T.~G. 2000, \jgr, 105, 2375
	
	\bibitem[{{Lin} {et~al.}(2003){Lin}, {Soon}, \& {Baliunas}}]{Lin2003}
	{Lin}, J., {Soon}, W., \& {Baliunas}, S.~L. 2003, \nar, 47, 53
	
	\bibitem[{{Liu}(2008)}]{Liu2008}
	{Liu}, Y. 2008, \apj, 679, L151
	
	\bibitem[{{Mari{\v{c}}i{\'c}} {et~al.}(2007){Mari{\v{c}}i{\'c}},
		{Vr{\v{s}}nak}, {Stanger}, {Veronig}, {Temmer}, \&
		{Ro{\v{s}}a}}]{Maricic2007}
	{Mari{\v{c}}i{\'c}}, D., {Vr{\v{s}}nak}, B., {Stanger}, A.~L., {Veronig},
	A.~M., {Temmer}, M., \& {Ro{\v{s}}a}, D. 2007, \solphys, 241, 99
	
	\bibitem[{{McCauley} {et~al.}(2015){McCauley}, {Su}, {Schanche}, \& {et
			al}}]{McCauley2015}
	{McCauley}, P.~I., {Su}, Y.~N., {Schanche}, N., \& {et al}. 2015, \solphys,
	290, 1703
	
	\bibitem[{{Moore} {et~al.}(2001){Moore}, {Sterling}, {Hudson}, \&
		{Lemen}}]{Moore2001}
	{Moore}, R.~L., {Sterling}, A.~C., {Hudson}, H.~S., \& {Lemen}, J.~R. 2001,
	\apj, 552, 833
	
	\bibitem[{{Neupert}(1968)}]{Neupert1968}
	{Neupert}, W.~M. 1968, \apjl, 153, L59
	
	\bibitem[{{Priest} \& {Forbes}(2002)}]{Priest2002}
	{Priest}, E.~R., \& {Forbes}, T.~G. 2002, \aapr, 10, 313
	
	\bibitem[{{Schou} {et~al.}(2012){Schou}, {Scherrer}, {Bush}, \& {et
			al}}]{Schou2012}
	{Schou}, J., {Scherrer}, P.~H., {Bush}, R.~I., \& {et al}. 2012, \solphys, 275,
	229
	
	\bibitem[{{Schrijver} \& {De Rosa}(2003)}]{Schrijver2003}
	{Schrijver}, C.~J., \& {De Rosa}, M.~L. 2003, \solphys, 212, 165
	
	\bibitem[{{Song} {et~al.}(2018{\natexlab{a}}){Song}, {Chen}, {Qiu}, {Chen}, \&
		{et al}}]{Song2018a}
	{Song}, H.~Q., {Chen}, Y., {Qiu}, J., {Chen}, C.~X., \& {et al}.
	2018{\natexlab{a}}, \apj, 857, L21
	
	\bibitem[{{Song} {et~al.}(2013){Song}, {Chen}, {Ye}, \& {et al}}]{Song2013}
	{Song}, H.~Q., {Chen}, Y., {Ye}, D.~D., \& {et al}. 2013, \apj, 773, 129
	
	\bibitem[{{Song} {et~al.}(2015){Song}, {Chen}, {Zhang}, {Cheng}, {Fu}, \&
		{LI}}]{Song2015}
	{Song}, H.~Q., {Chen}, Y., {Zhang}, J., {Cheng}, X., {Fu}, H., \& {LI}, G.
	2015, \apjl, 804, L38
	
	\bibitem[{{Song} {et~al.}(2018{\natexlab{b}}){Song}, {Zhou}, {Li}, {Cheng},
		{Zhang}, {Chen}, {Chen}, {Ma}, {Wang}, \& {Zheng}}]{Song2018b}
	{Song}, H.~Q., {Zhou}, Z.~J., {Li}, L.~P., {Cheng}, X., {Zhang}, J., {Chen},
	Y., {Chen}, C.~X., {Ma}, X.~W., {Wang}, B., \& {Zheng}, R.~S.
	2018{\natexlab{b}}, \apjl, 864, L37
	
	\bibitem[{{Sterling} \& {Moore}(2005)}]{Sterling2005}
	{Sterling}, A.~C., \& {Moore}, R.~L. 2005, \apj, 630, 1148
	
	\bibitem[{{Sun}(2018)}]{Sun2018}
	{Sun}, X. 2018, arXiv e-prints, arXiv:1801.04265
	
	\bibitem[{{Temmer} {et~al.}(2010){Temmer}, {Veronig}, {Kontar}, {Krucker}, \&
		{Vr{\v s}nak}}]{Temmer2010}
	{Temmer}, M., {Veronig}, A.~M., {Kontar}, E.~P., {Krucker}, S., \& {Vr{\v
			s}nak}, B. 2010, \apj, 712, 1410
	
	\bibitem[{{Thompson}(2009)}]{Thompson2009}
	{Thompson}, W.~T. 2009, \icarus, 200, 351
	
	\bibitem[{{Titov} \& {D{\'e}moulin}(1999)}]{Titov1999}
	{Titov}, V.~S., \& {D{\'e}moulin}, P. 1999, \aap, 351, 707
	
	\bibitem[{{T{\"o}r{\"o}k} \& {Kliem}(2005)}]{Torok2005}
	{T{\"o}r{\"o}k}, T., \& {Kliem}, B. 2005, \apjl, 630, L97
	
	\bibitem[{{T{\"o}r{\"o}k} \& {Kliem}(2007)}]{Torok2007}
	---. 2007, Astronomische Nachrichten, 328, 743
	
	\bibitem[{{van Ballegooijen} \& {Martens}(1989)}]{VanBallegooijen1989}
	{van Ballegooijen}, A.~A., \& {Martens}, P.~C.~H. 1989, \apj, 343, 971
	
	\bibitem[{{Vasantharaju} {et~al.}(2018){Vasantharaju}, {Vemareddy}, {Ravindra},
		\& {Doddamani}}]{Vasantharaju2018}
	{Vasantharaju}, N., {Vemareddy}, P., {Ravindra}, B., \& {Doddamani}, V.~H.
	2018, \apj, 860, 58
	
	\bibitem[{{Vemareddy} \& {Dem{\'o}ulin}(2018)}]{Vemareddy2018}
	{Vemareddy}, P., \& {Dem{\'o}ulin}, P. 2018, \apj, 857, 90
	
	\bibitem[{{Vemareddy} {et~al.}(2017){Vemareddy}, {Gopalswamy}, \&
		{Ravindra}}]{Vemareddy2017}
	{Vemareddy}, P., {Gopalswamy}, N., \& {Ravindra}, B. 2017, \apj, 850, 38
	
	\bibitem[{{Vemareddy} \& {Zhang}(2014)}]{Vemareddy2014}
	{Vemareddy}, P., \& {Zhang}, J. 2014, \apj, 797, 80
	
	\bibitem[{{Vourlidas}(2014)}]{Vourlidas2014}
	{Vourlidas}, A. 2014, Plasma Physics and Controlled Fusion, 56, 064001
	
	\bibitem[{{Vr{\v s}nak}(2016)}]{Vrsnak2016}
	{Vr{\v s}nak}, B. 2016, Astronomische Nachrichten, 337, 1002
	
	\bibitem[{{Wuelser} {et~al.}(2004){Wuelser}, {Lemen}, {Tarbell}, \& {et
			al}}]{Wuelser2004}
	{Wuelser}, J.-P., {Lemen}, J.~R., {Tarbell}, T.~D., \& {et al}. 2004, in
	Society of Photo-Optical Instrumentation Engineers (SPIE) Conference Series,
	Vol. 5171, Telescopes and Instrumentation for Solar Astrophysics, ed.
	S.~{Fineschi} \& M.~A. {Gummin}, 111--122
	
	\bibitem[{{Xie} {et~al.}(2013){Xie}, {Gopalswamy}, \& {St. Cyr}}]{Xie2013}
	{Xie}, H., {Gopalswamy}, N., \& {St. Cyr}, O.~C. 2013, \solphys, 284, 47
	
	\bibitem[{{Zhou} {et~al.}(2019){Zhou}, {Cheng}, {Zhang}, {Wang}, {Wang}, {Liu},
		{Zhuang}, \& {Cui}}]{Zhou2019}
	{Zhou}, Z., {Cheng}, X., {Zhang}, J., {Wang}, Y., {Wang}, D., {Liu}, L.,
	{Zhuang}, B., \& {Cui}, J. 2019, \apjl, 877, L28
	
	\bibitem[{{Zuccarello} {et~al.}(2015){Zuccarello}, {Aulanier}, \&
		{Gilchrist}}]{Zuccarello2015}
	{Zuccarello}, F.~P., {Aulanier}, G., \& {Gilchrist}, S.~A. 2015, \apj, 814, 126
	
	\bibitem[{{Zuccarello} {et~al.}(2016){Zuccarello}, {Aulanier}, \&
		{Gilchrist}}]{Zuccarello2016}
	---. 2016, \apj, 821, L23
	
	\bibitem[{{Zuccarello} {et~al.}(2014){Zuccarello}, {Seaton}, {Filippov},
		{Mierla}, \& {et al}}]{Zuccarello2014}
	{Zuccarello}, F.~P., {Seaton}, D.~B., {Filippov}, B., {Mierla}, M., \& {et al}.
	2014, \apj, 795, 175
	
\end{thebibliography}


\end{document}